# A study of the kinematics near WR-stars in the IC10 galaxy

Podorvanyuk N.Yu. (nicola@sai.msu.ru)

Sternberg Astronomical Institute, Moscow State University, Moscow, Russia

## 1. Introduction

Among the other dwarf galaxies with violent star formation IC10 galaxy stands out for anomalously high number of Wolf-Rayet (WR) stars. For instance, the number of WR-stars in IC10 is 20 times higher than that in LMC and spatial density of WR stars in IC10 reaches 11 WR stars per square kpc (Massey et al., 2002 [8], Massey et al., 2007 [7], Crowther et al., 2003 [4] and references in these papers). This is the highest density of WR stars among the dwarf galaxies, comparable to that of massive spiral galaxies.

Provided the regular initial mass function (IMF), such high WR star density presumes almost "simultaneous" outbreak of modern star formation, sweeping most part of the galaxy.

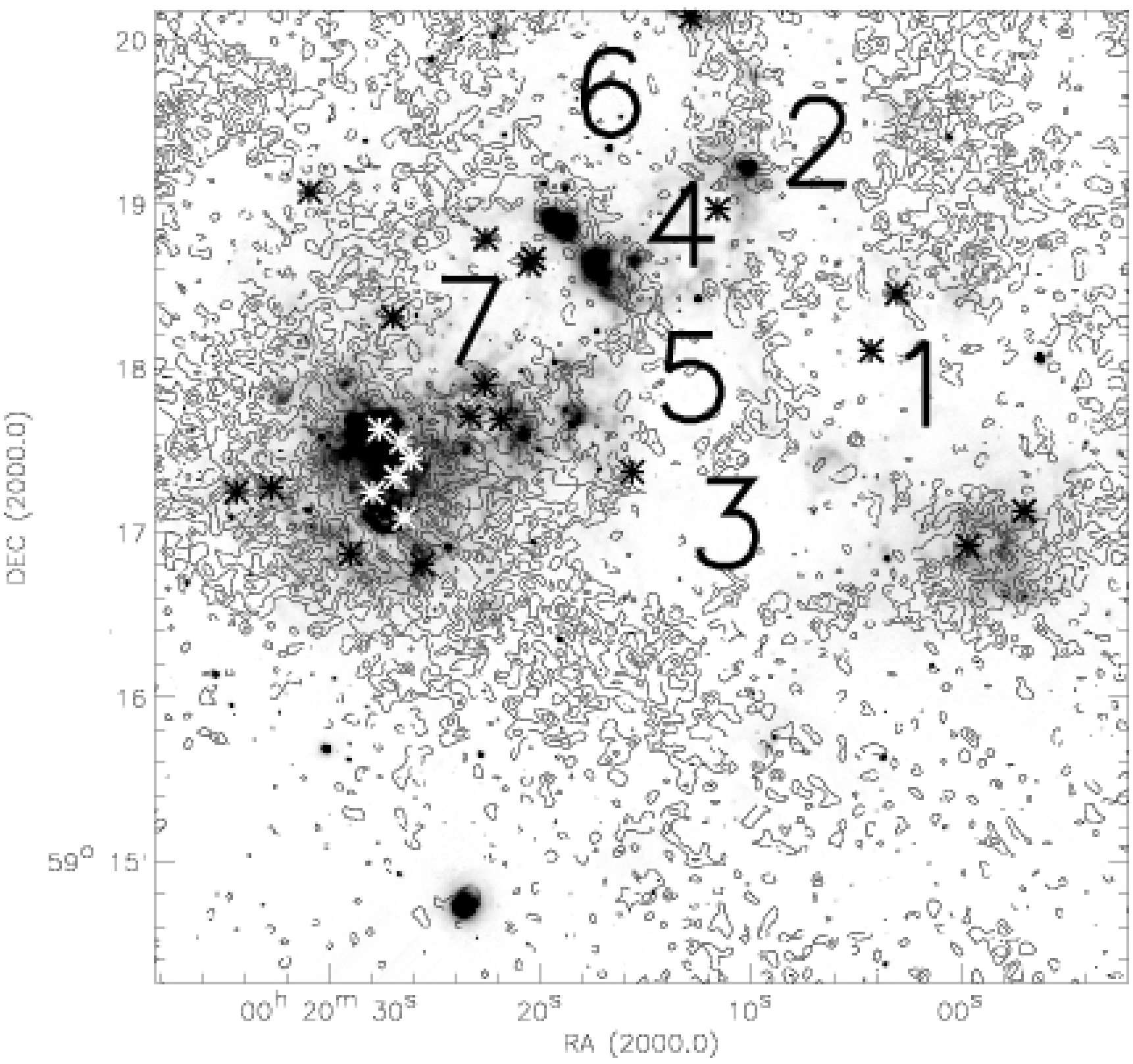

**Fig. 1.** Wolf-Rayet stars (*), plotted against the Hα image of the central part of IC10 galaxy. Contours show the map of 21cm intensity distribution. Numbers mark the envelopes of the neutral gas as given in Wilcots & Miller paper.

On the Hα image of the galaxy one can distinguish many arc-shaped structures of different scale and complex morphology around the Wolf-Rayet stars (Fig. 1). No clear and closed shells can be seen around any of the WR stars which can be easily explained by the fact that the whole galaxy is actually one common complex of modern star formation with plenty of star wind sources. Star wind of multiple stars experiences interference which results in formation of complex structures and kinematics.

In current paper we study the kinematics of neutral and ionized gas in the vicinity of Wolf-Rayet stars with the goal of finding its connection to the enveloping structures and determining the expansion effect on these structures. Spectrascopically approved in Royer et al. 2001 [10] Wolf-Rayet stars are marked as "R", WR stars from Massey & Holmes, 2002 are marked as "M".

The author of this paper has plotted the "position-velocity" diagrams (PV-diagrams) along several sections around every WR star. Each section plotted in ionized gas line was doubled for 21cm observations as well. The width of the sections varied from 0,5" to 5". In PV-diagrams listed in the current paper the width is 2" for ionized and 4" for neutral gas.

For precise measurement of velocities where it was necessary the Voigt profile (the convolution of Gaussian and Lorentz profiles) was fitted into the certain section element on the PV-diagram and the velocity was determined according to the position of profile maximum.

For the interpretation of the results acquired based on 21cm observations the map of column density distribution of neutral hydrogen was plotted for the studied region of IC10 galaxy. From the classic theory the radiative stage of the interaction between the star wind and the interstellar medium for a separate star can be described by the system of equations of motion, one of which gives the velocity of shell v in km/s provided the power of star wind $L_w$, density of the unperturbed medium n and the age of the star t:

$$v\ (t) = 16(L_{36} / n)^{1/5}\ (t_6)^{-2/5}\ km/s,$$

where $L_{36}=L_w/10^{36}$ erg/s, $t_6=t/10^6$ years.

From our observations in Hα we can determine the expansion velocity of the shell or its part and the shell radius. The relation of radius to velocity gives an estimate of the star age. We can determine the special density for the specific area of neutral gas by dividing the column density by the size of the emitting region. Thus, form the equation given above we can determine the power of the wind, as the expansion velocity, density of the medium and star age are already known from the observations.

# 2. Observations and data reduction

### 2.1 Interferometric Hα observations on the 6-metertelescope SAO RAS

Interferometric observations of IC 10 were performed on September 8/9, 2005, at the prime focus of the 6-m telescope using the SCORPIO focal reducer; the equivalent focal ratio of the system was F/2.6. SCORPIO was described by Afanasiev and Moiseev (2005) and on the Internet (http://www.sao.ru/hq/lsfvo/devices.html); the SCOPRIO capabilities in interferometric observations were also described by Moiseev (2002) [9]. We used a scanning Fabry–Perot interferometer (FPI) operating in the 501st order at the Hα wavelength. The spacing between the neighboring orders of interference, Δλ=13Å, corresponded to a region free from order overlapping of ~ 600 km/s on the radial velocity scale. The FWHM of the instrumental profile was ~ 0.8Å, or ~35 km/s. Premonochromatization was performed using an interference filter with FWHM=13Å centered on the Hα line. The detector was an EEV 42–40 2048×2048-pixel CCD array. The observations were performed with 2×2-pixel binning to reduce the readout time. In each spectral channel, we obtained 1024×1024-pixel images at a scale of 0.35″ per pixel; the total field of view was 6.1′.
During our observations, we successively took 36 interferograms of the object for various FPI plate spacings, so the width of the spectral channel corresponded to Δλ=0.37Å, or 17 km/s near Hα. To properly subtract parasitic ghosts from the galaxy's numerous emission regions, the observations were performed for two different orientations of the instrument's field of view. The total exposure time was 10 800 s; the seeing (the FWHM of field-star images) varied within the range 0.8"-1.3".
We reduced the observations using software running in the IDL environment (Moiseev 2002) [9]. After the primary reduction, the observational data were represented as 1024×1024×36 data cubes; here, a 36-channel spectrum corresponds to each pixel. The final angular resolution (after smoothing during the data reduction) was ∼1.2". The formal accuracy of the constructed wavelength scale was about 3–5 km/s.

### 2.2 21cm observations and data reduction

Kinematical study of the neutral hydrogen in the galaxy in this paper is based on the 21cm VLA data obtained by Wilcots & Miller [2] and kindly provided by the authors for re-analysis.
The data used in this paper is a combination of observational data obtained on VLA in B,C and D configurations during the period of 1993-1994. The channel width corresponds to 2.57 km/s. Standard calibration techniques and mapping were carried out in AIPS package. The image size for each spectral channel was 1024x1024 pixels with pixel size of 1.5". The data reduction results are presented as a data cube of 1024x1024x66 elements with the angular resolution of 4".7x5".0, which corresponds to linear resolution ~19 parsec provided the distance to the galaxy is 790 kiloparsec.
The range of velocities covered by the 21cm observations is -419 km/s ... -252 km/s.

## 3. Kinematics of the near Wolf-Rayet stars

Here we will demonstrate the kinematical study of the Wolf-Rayet star vicinity using two stars M1 and M2 as an example. On the Hα plot of the vicinities of these stars (Figure 2) one can clearly see the characteristic arc-shaped structures, which can be the "walls" of the star envelopes blown away by the star wind. The M2 star is located inside the HI cloud (borders to shells 1 and 3 according to [2]), the M1 star is on the edge of this cloud. The distance between M2 and its alleged shell "wall" varies from 5 to 15 arc seconds (20-60 parsec). The distance from M1 to its "wall" lies between 5 and 10 arc seconds (20-40 parsec).
The mean velocity of the unperturbed gas in this part of the galaxy according to our FPI data is -320±10 km/s. However, the observed velocities in the close vicinity of M1 and M2 are quite different from this mean value (Figure 2). The section on this figure successively goes through M1 and M2. In the small distinguishable cloud of ionized gas located several seconds to the South-East of M2, PV-diagram (10 arc seconds) shows the mean velocity of -348 km/s. In the close vicinity of M2 (18 arc seconds) the velocity of the ionized gas is -351 км/с. Between the star and visible "wall" of the shell (23-27 arc seconds) the velocity rises to -340 km/s, and in the region on the "wall" itself (30 arc seconds) it drops to -355 km/s.

There is another structure located approximately 5 arc seconds from the “wall” which is likely a part of the shell blown away by M1 star. In the vicinity of this “wall” the velocity is approximately -340 km/s. The same velocity is observed in the vicinity of M1 itself (42 arc seconds). 5 arc seconds away from the stars the intensity of the gas drops down and its velocity is -345 km/s. Several arc seconds further the velocity experiences rapid change to the value of -315 km/s. Only at the very edge of the PV-diagram the velocity drops again to the value of -342 km/s.

Kinematics of the neutral gas in this particular region of the galaxy shows that the cloud of neutral gas where M2 is located and at the edge of which is M1 shows the velocity of -360km/s on its border (which differs by 50 km/s from the velocity of unperturbed gas!)

The central part is approaching at the velocity of approximately 15-20 km/s, and the maximum velocity on the PV-diagram is observed notably in the same region as the star.

The results can be interpreted rather deliberately: the stars M1 and M2 in the galaxy demonstrate the vivid example of the star wind influence on the interstellar medium.

M1, being located on the edge of the neutral gas cloud, has created the part of the inflated shell by its star wind. The velocity of the shell expansion is approximately 25 km/s compared to the velocity of unperturbed part of the galaxy. In the opposite direction where the neutral gas density in the close vicinity of M1 is low, the “void” being “sweeped” away by the star is observed (emission comes only from the unperturbed gas from the other parts of the galaxy along the line of sight). Where the density of the neutral gas increases again (the edge of the cloud), almost the same velocities as on the “wall” are observed.

The estimate of the Wolf-Rayet star M1 wind power, made using the technique described above gives the value of $L=53\cdot10^{36}$ erg/s.

M2 star, located inside the neutral gas cloud also has no evident closed swept away shell, but this can be easily explained by the density distribution of the neutral hydrogen in its vicinity. The density to the South of the star is very high, a little lower to the West of it and even lower and decreasing in the Northern and Eastern directions. Thus, compared to the mean velocity of the unperturbed gas, the velocity of the expansion of the part of the M2 shell depending on the direction lies between 20-35 km/s.

Using the velocities in two opposite directions from the star based on the PV-diagram, the star wind power of M2 can be found. In the South-Eastern direction, where the “wall” is formed by the dense HI cloud with the density of $7cm^{-3}$, the star power is $L=74\cdot10^{36}$ erg/s. In the direction of M1 where the “wall” is located on the distance of approximately 45 parsec, $L=96\cdot10^{36}$ erg/s. When estimating the wind power in the second case it was taken into consideration that the wind from the star had passed the low-density region of neutral hydrogen first (approximately $1cm^{-3}$) and then had been slowed down by the high-density (approximately $7cm^{-3}$) medium.

During the work the author of the paper has conducted a study of 23 Wolf-Rayet star vicinities, which can be divided into three groups. The first group is the stars that showed impact on the surrounding medium. Those are M1, M2 and 6 other syars: R6, R13, M7, M10, R11, R12. In general these are the stars which are located far from the other star wind sources.

The vicinities of these stars were studied as described above for M1 and M2. For the most of the stars the kinematical evidence of star wind influence on the interstellar medium was found and the estimates of the star wind power were done.

The second group is the stars located in the star formation region (the brightest region on Hα map of IC10 galaxy). It is rather hard to identify the influence of the specific star on the interstellar medium because there are lots of star wind sources, for instance several young star clusters, in this complex alongside with the studied WR stars. Because of the interference of the star wind coming from various sources, no evident shells around the WR stars are observed and only the local broadenings can be seen.

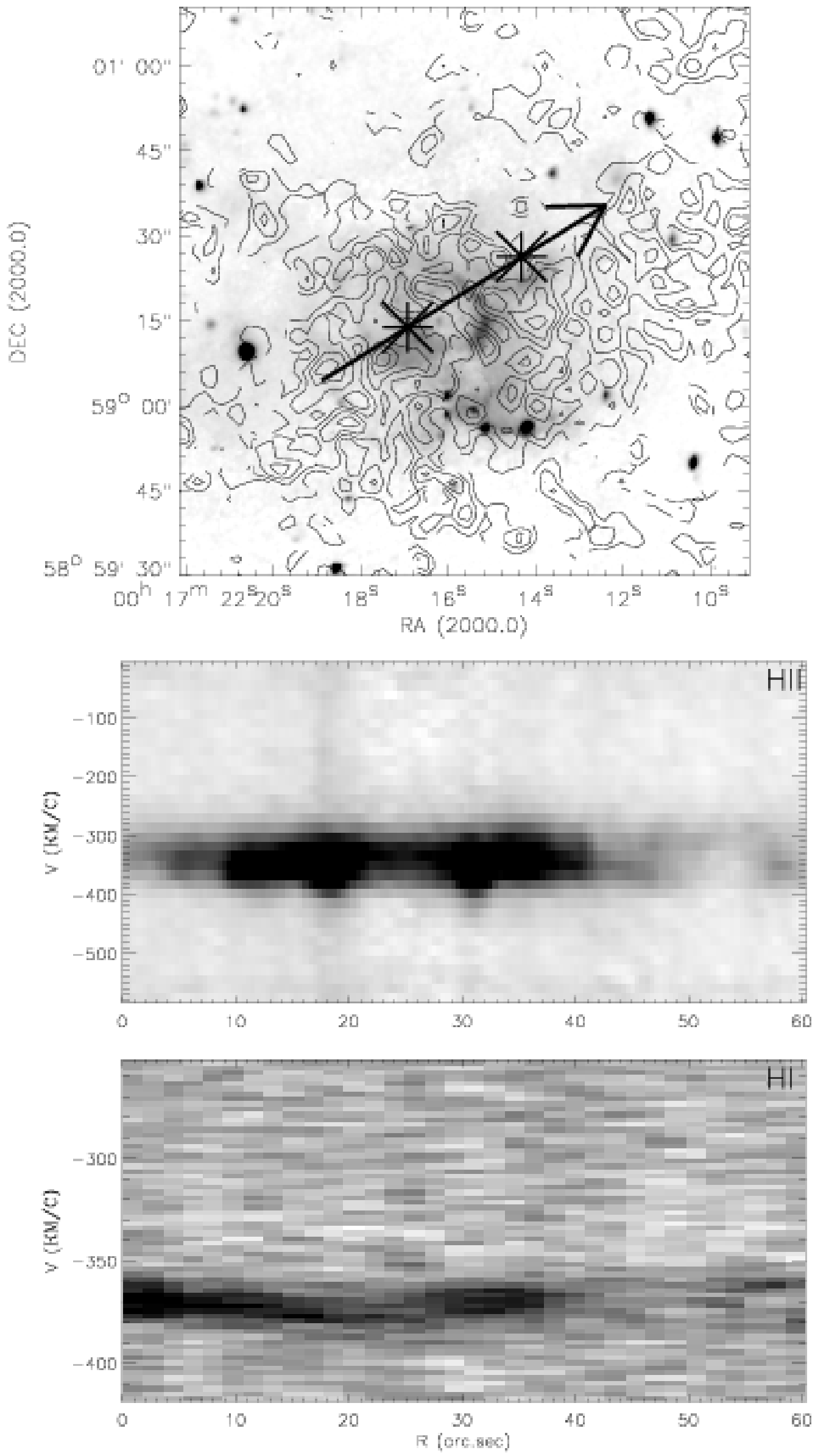

**Fig. 2.** Position-velocity diagram ploted through the section across M1 (42 arc seconds) and M2 (18 arc second). Hα and 21cm observations.

Practically along all of the sections used for constructing PV-diagrams, the velocities of the main Hα component are the same in the vicinities of all the stars and have the value of approximately -320 km/s. The second group includes 8 stars: M12, M13, M14, R10, M24, R2, M23, M20. The estimates of the star wind power have been made based on the observed signs of the star wind influence on the interstellar medium.

Seven Wolf-Rayet stars which show no star wind influence on the interstellar medium -- M4, M5, R8, R9, M9, M15 and R5 were included into the third group.

The estimates of the star wind power for Wolf-Rayet stars obtained in this paper are given in Table 1.

## 4. Discussion

The vicinities of 23 Wolf-Rayet stars were studied in the paper. For 8 stars the evident signs of star wind influence on the interstellar medium were found. In general, these stars are located far from the other sources of the star wind. Due to the heterogeneity of the environment none of the stars demonstrate the clear closed shell ring, however some “walls”, arcs and the other structures are observed, which revealed velocity (from 5 to 50 km/s) and could be the result of a Wolf-Rayet stars influencing the interstellar medium. Eight more stars are located in the complex of modern star formation of IC10 galaxy. Since all these objects are located very close to each other and there are several young star clusters (which are the strong star wind sources themselves) together with the Wolf-Rayet stars it is hard to find any evident signs of the WR influence on the surrounding environment in this part of the galaxy. However, we managed to find the velocities of about 5…15km/s in the surrounding "walls" for six stars out of eight.

| **WR star** | **Right Ascension α** | **Declination δ** | **Star power, *L* (erg/s)** |
|---|---|---|---|
| Stars that have shown the influence on the interstellar medium | | | |
| M1 | $00^h19^m56^s.97$ | +59°17'08".0 | $53{\cdot}10^{36}$ |
| M2 | $00^h19^m58^s.65$ | +59°16'55".3 | $74\ldots96{\cdot}10^{36}$ |
| R6 | $00^h20^m03^s.02$ | +59°18'27".4 | $7{\cdot}10^{36}$ |
| R13 | $00^h20^m15^s.62$ | +59°17'22".2 | $7{\cdot}10^{36}$ |
| M7 | $00^h20^m21^s.87$ | +59°17'41".5 | $4{\cdot}10^{36}$ |
| M10 | $00^h20^m23^s.36$ | +59°17'42".6 | $25{\cdot}10^{36}$ |
| R11 | $00^h20^m22^s.68$ | +59°17'53".9 | $18{\cdot}10^{36}$ |
| R12 | $00^h20^m25^s.61$ | +59°16'48".6 | $37{\cdot}10^{36}$ |
| Stars that are located in the star formation complex | | | |
| M12 | $00^h20^m26^s.17$ | +59°17'26".9 | $4{\cdot}10^{36}$ |
| M13 | $00^h20^m26^s.63$ | +59°17'33".2 | $4{\cdot}10^{36}$ |
| M14 | $00^h20^m26^s.87$ | +59°17'20".2 | $4{\cdot}10^{36}$ |
| R10 | $00^h20^m26^s.48$ | +59°17'05".3 | $8{\cdot}10^{36}$ |
| M24 | $00^h20^m27^s.67$ | +59°17'37".7 | $20{\cdot}10^{36}$ |
| R2 | $00^h20^m28^s.00$ | +59°17'37".7 | $12{\cdot}10^{36}$ |
| M23 | $00^h20^m32^s.79$ | +59°17'16".4 | $1{\cdot}10^{36}$ |
| M20 | $00^h20^m34^s.46$ | +59°17'14".7 | $1{\cdot}10^{36}$ |
| Stars that haven’t shown any influence on the interstellar medium | | | |
| M4 | $00^h20^m11^s.55$ | +59°18'58".3 | |
| M5 | $00^h20^m12^s.85$ | +59°20'08".5 | |
| R8 | $00^h20^m20^s.56$ | +59°18'37".8 | |
| R9 | $00^h20^m20^s.33$ | +59°18'40".2 | |
| M9 | $00^h20^m22^s.60$ | +59°18'47".3 | |
| M15 | $00^h20^m27^s.03$ | +59°18'18".6 | |
| R5 | $00^h20^m04^s.24$ | +59°18'06".6 | |

**Table 1.** Estimate of the Wolf-Rayet star power in IC10 galaxy

For the stars that have prominent alleged inflated shell parts the estimate of the star wind power has been done. It was shown, that the wind power needed for the formation of the observed structures with the observed expansion velocity is approximately ~(0.01-0.84) $10^{38}$ erg/s in the medium with unperturbed density of 1…10 $cm^{-3}$. Note that the calculations are rough and one can only estimate the order of magnitude of the star wind power. Nevertheless, the obtained values of the star wind power are characteristic ones for the Wolf-Rayet stars even considering the low metallicity of the galaxy, which can reduce the star wind power by the factor of 2 or 3.
Seven stars haven't shown any influence o the interstellar medium.
The absence of the prominent shell around the Wold-Rayet stars can be due to the low density of the surrounding gas, being swept away by the star on the previous stages or as a result of simultaneous influence of multiple star associations. In this case the ring nebulae can be observed only if the Wolf-Rayet star wind is still sweeping the matter, ejected by the star. As it was shown in the paper by Dopita, Lozinskaya et al. [3], the characteristic size of such ring nebulae is not larger than 4-7 parsec. It is practically impossible to detect the shell of such size on the images used in this paper, because 4-7 parsec correspond to 3-5 pixels for Hα observations on FPI on the 6-meter SAO RAS telescope, and 1 pixel for HI observations.

# 5. Conclusion

In the paper 23 Wolf-rayet star vicinities were studied. The detailed study of kinematics of the ionized and neutral gas in the vicinity of Wolf-Rayet stars in IC10 galaxy was done, resulting in detection of star wind influence on the interstellar medium for most of the studied stars. The estimates of star wind power were done for the stars showing the evident signs of their influence on the surrounding gas. In general, for the studied region of the galaxy the Wolf-rayet star power is about ~(0.01-0.84) $10^{38}$ erg/s, which is characteristic Wolf-Rayet star power even considering the low metallicity of the galaxy.
The studies of seven stars haven't shown prominent signs of their influence on the interstellar medium. This is likely due to low density of surrounding gas, being swept away by the star on the previous stages or as a result of simultaneous influence of many stars of associations. It was shown that the ring nebulae, which should be observed in this case, cannot be resolved in the data we have.
The studies were carried out with the financial support of the Russian Foundation for Basic Research (project 07--02--00227). The work is based on the observational material obtained using the 6-meter SAO RAS telescope funded by the Ministry of Science of Russian Federation (registration number 01-43). The author thanks T.A.Lozinskaya, A.V.Moiseev and O.K.Silchenko for the useful advices, comments and miscellaneous help.